**Piotr GAS**

AGH University of Science and Technology

# Temperature inside tumor as time function in RF hyperthermia

***Abstract***. *A simplified 2-D model which is an example of regional RF hyperthermia is presented. Human body is inside the wire with exciting current and the electromagnetic energy is concentrated within the tumor. The analyzed model is therefore a coupling of the electromagnetic field and the temperature field. Exciting current density in human body has been calculated using the finite element method, and then bioheat equation in time-depended nonstationary case has been resolved. At the and obtained results are presented.*

***Streszczenie***. *W niniejszej pracy przedstawiono uproszczony model dwuwymiarowy stanowiący prosty przykład zastosowania regionalnej hipertermii o częstotliwości radiowej, w której ciało człowieka znajduje się wewnątrz przewód z wymuszającym prądem, a energia elektromagnetyczna zostaje skupiona w środku guza. Analizowany model stanowi zatem sprzężenie pola elektromagnetycznego i pola temperatury. Posługując się metodą elementów skończonych na wstępie wyznaczono gęstość prądu indukowanego w ciele człowieka, a następnie rozwiązano biologiczne równanie ciepła w przypadku niestacjonarnym zależnym od czasu. Na końcu zestawiono uzyskane wyniki.* **(Temperatura wewnątrz guza jako funkcja czasu w RF hipertermii)**

**Słowa kluczowe**: hipertermia, biologiczne równanie ciepła, rozkład temperatury, metoda elementów skończonych

## Introduction

The rapid technological development and growth of knowledge about the biological effects of electromagnetic fields have become the starting point for a wide range of biomedical applications. In addition to medical diagnosis, they are focused primarily on therapeutic applications. Medical use of electromagnetic fields is reflected inter alia in hyperthermic oncology in the treatment of cancer.

Hyperthermia, also known as thermotherapy uses a well-known phenomenon of induced temperature increase inside the tissue exposed to the action of electromagnetic energy of radio and microwave frequencies. When the temperature rises slightly above the normal human body temperature of 37°C, it is fully controlled by the thermoregulation mechanisms of the body. However, when growth is higher, it can cause irreversible changes in the biological proteins including denaturation [9]. The studies clearly show that electromagnetic heating of tumors to temperatures in the range of 40 – 44°C can lead to partial or complete destruction of tumor cells while not causing negative changes in the healthy tissue surrounding the tumor [8, 18]. In addition, evidence shows that the effectiveness of hyperthermia increases significantly in combination with other cancer treatments such as radiotherapy or chemotherapy [4, 7, 10, 11, 14].

Apparently, an extremely important issue is to control the temperature distribution in the treated area to avoid excessive temperature increase in the normal tissues surrounding the tumor [1]. There are many studies on the treatment of cancer using hyperthermia which demonstrates that this aspect is still important and more research is needed in this matter [2, 5, 6, 13].

## Main equations

Let us consider a cross section of the human body as shown in Figures 1 and 2. The human body is approximated to an ellipse whose semi-axes are respectively $a = 20$ cm and $b = 12$ cm. Inside the body there is a tumor with the radius of $r_1 = 2.5$ cm. An elliptical wire with the semi-axes $A = 50$ cm and $B = 40$ cm is placed around the human body. Through the wire an alternating current flows in clock-wise direction with the amplitude $I_m = 16$ A and frequency $f = 100$ MHz. The exciting current generates a sinusoidal electromagnetic field, which next induces eddy currents in human body. Eddy currents are a source of heat and after transient time a temperature distribution in human body is established Therefore, in the analyzed model, we deal with the electromagnetic field coupled with the field temperature.

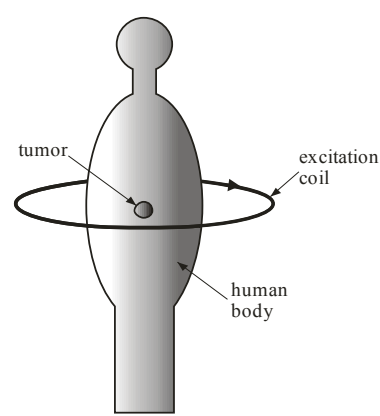

Fig.1. Schematic view of human body surrounded by wire with excitation current

Let us start with Maxwell's equations in the frequency domain:

(1) $$\nabla \times \mathbf{H} = \mathbf{J}_i + \mathbf{J}_c + j\omega \mathbf{D}$$

(2) $$\nabla \times \mathbf{E} = -j\omega \mathbf{B}$$

where **E** and **H** are respectively the electric and magnetic field strengths, $\mathbf{J}_i$ is an impressed current density, which is treated as a source of electromagnetic field, $\mathbf{J}_c$ is conduction current resulting from the existence of an electric field according to Ohm's law

(3) $$\mathbf{J}_c = \sigma \mathbf{E}$$

where $\sigma$ is the electrical conductivity of the body. **D** and **B** are respectively the vectors of electric displacement density and magnetic induction given in the form of material dependences:

(4) $$\mathbf{D} = \varepsilon \mathbf{E}$$

(5) $$\mathbf{B} = \mu \mathbf{H}$$

where $\varepsilon$ and $\mu$ are respectively the permittivity and magnetic permeability of the medium.

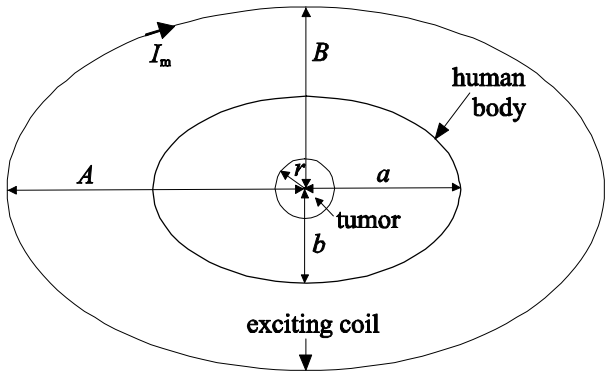

Fig.2. Cross section of the human body with a tumor inside and exciting wire together with geometrical dimensions

After the introduction of the magnetic vector potential

(6) $$\mathbf{B} = \nabla \times \mathbf{A}$$

one can derive the following equation describing the field distribution:

(7) $$\nabla \times \left(\frac{1}{\mu} \nabla \times \mathbf{A}\right) + \left(j\omega\sigma - \omega^2 \varepsilon\right)\mathbf{A} = \mathbf{J}_i$$

Since the magnetic vector potential in the time domain $\mathbf{A}(\mathbf{r}, t)$ is related with complex amplitude by

(8) $$\mathbf{A}(\mathbf{r},t) = \text{Re}\left[\hat{\mathbf{A}}(\mathbf{r}) e^{j\omega t}\right]$$

therefore equation (7) in the complex domain is given by

(9) $$\nabla \times \left(\frac{1}{\mu} \nabla \times \hat{\mathbf{A}}\right) + \left(j\omega\sigma - \omega^2 \hat{\varepsilon}\right)\hat{\mathbf{A}} = \hat{\mathbf{J}}_i$$

where $\hat{\varepsilon}$ is the complex permittivity defined as

(10) $$\hat{\varepsilon} = \varepsilon - j\frac{\sigma}{\omega}$$

The total induced current density in the human body is calculated from the equation

(11) $$\hat{\mathbf{J}} = \sigma \hat{\mathbf{E}} = -j\omega\sigma \hat{\mathbf{A}}$$

The second basic equation used in the presented simulation is the so-called bioheat equation given by Pennes in the mid-twentieth century [12, 15, 17]. It describes the phenomenon of transport and heat transfer in biological tissues. It is assumed that heat exchange with blood takes place only through the capillary perfusion, neglecting heat transfer derived from the larger blood vessels. This simplification is fully sufficient for the analysis of the situation of heat exposure. In transient analysis the bioheat equation is expressed by

(12) $$\rho C \frac{\partial T}{\partial t} + \nabla \cdot (-k\nabla T) = \rho_b C_b \omega_b (T_b - T) + Q_{ext} + Q_{met}$$

where $T$ is body temperature [K], $k$ − tissue thermal conductivity [W/(m² K)], $\rho$ − tissue density [kg/m³], $C$ − tissue blood specific heat [J/(kg K)], $T_b$ − blood vessel temperature [K], $\rho_b$ − blood density [kg/m³], $\omega_b$ − blood perfusion rate [1/s], $C_b$ − blood specific heat [J/(kg K)].

The described model also takes into account both the metabolic heat generation rate $Q_{met}$ [W/m³] as well as the external heat sources $Q_{ext}$ [W/m³], which is responsible for the changing of the temperature inside the exposed body according to the following equation

(13) $$Q_{ext} = \sigma \hat{\mathbf{E}} \cdot \hat{\mathbf{E}}^* = \sigma \left|\hat{\mathbf{E}}\right|^2$$

The bioheat equation allows us to assess both the transient response and the steady-state of temperature changes in the human body. Since this is a differential equation in time and space its solution requires to specify both the initial and boundary conditions to be specified. The initial condition is obtained as the solution of equation (12) in a steady state in the absence of external energy sources ($Q_{ext} = 0$), which corresponds to the physiological parameters of the body at temperature of

(14) $$T = T_0 = 37°C$$

The boundary condition explains heat exchange between the surface of the body and the external environment according to the equation [3, 16] given by

(15) $$\mathbf{n} \cdot (-k\nabla T) = h(T_{air} - T)$$

where $h$ is the heat transfer coefficient [W/(m²·K)], $T_{air}$ is the temperature of the air surrounding the body [K] and $\mathbf{n}$ is the unit vector normal to the surface. The term on the right side of above equation describes the heat losses due to convection, therefore a constant $h$ is also named as the convection coefficient.

**Simulation results**

In the analyzed model, the human body and tumor are considered as homogeneous media with averaged material parameters, therefore the simulation results may differ from those obtained in reality during hyperthermia treatment. The physical parameters of the model are given in Tables 1 and 2. In addition, heat transfer coefficient was assumed equal to $h = 10$ [W/(m²·K)], and the temperature of the air surrounding the human body equal to $T_{air} = 293.15$ [K], which corresponds to room temperature of 20°C.

Table 1. Physical parameters of tissues used in the numerical model

| Tissue | $\varepsilon_r$ | $\sigma$ [S/m] | $k$ [W/(m·K)] |
|---|---|---|---|
| Human body | 29.6 | 0.053 | 0.22 |
| Tumor | 160 | 0.64 | 0.56 |

| Tissue | $\rho$ [kg/m³] | $C$ [J/(kg K)] | $Q_{met}$ [W/m³] |
|---|---|---|---|
| Human body | 1050 | 3700 | 300 |
| Tumor | 1050 | 3700 | 480 |

Table 2. Physical parameters of blood taken in the bioheat equation

| Tissue | $\rho_b$ [kg/m³] | $C_b$ [J/(kg·K)] | $T_b$ [K] | $\omega_b$ [1/s] |
|---|---|---|---|---|
| Blood | 1020 | 3640 | 310.15 | in body 0.005<br>in tumor 0.0004 |

Equations (9) and (12) with appropriate initial and boundary conditions were solved using the finite element method. The simulation results are summarized in Figures 3 – 9. Fig. 3 shows the equipotential lines of the module of the magnetic vector potential. This vector is perpendicular to the x-y plane and its maximum value is close to the wire with the exciting current.

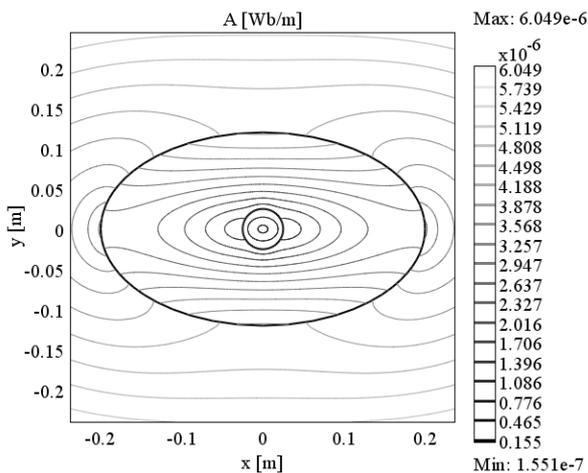

Fig.3. Equipotential lines of the modulus of magnetic vector potential

Equipotential lines of the vector current density induced in the human body is presented in Fig. 4. As expected, the highest value occurs within the tumor.

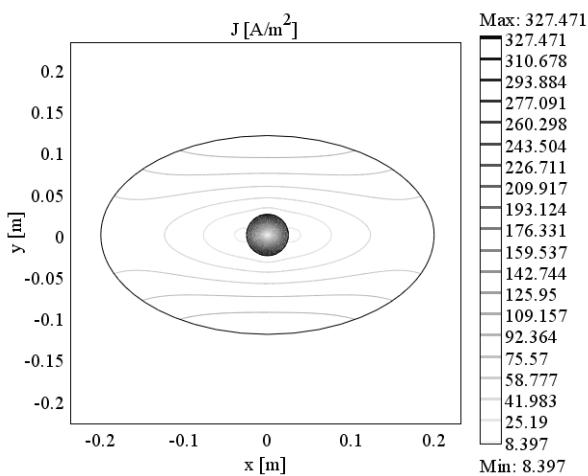

Fig.4. Equipotential lines of current density induced in the human body

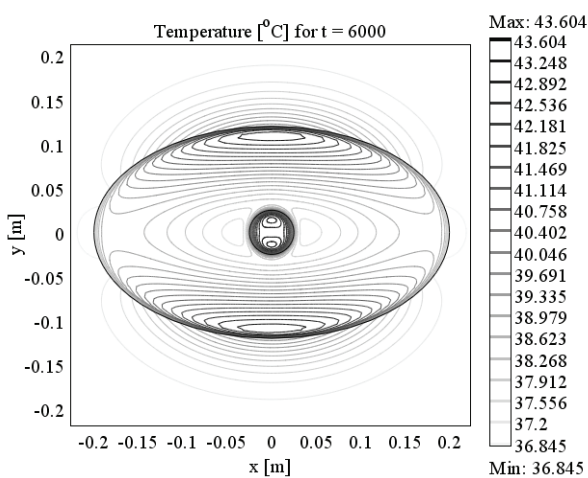

Fig.5. Isotherms in the human body for the steady state analysis

The rest drawings illustrate the temperature dependence within the tumor in different form. Fig. 5 represents the distribution of isotherms in the analysed model for the steady state after a time of $t = 6000$ [s].

Temperature distribution along horizontal and vertical symmetry axes of the human body for different moments of time, ranging from 0 to 6000 seconds, are presented in Figures 6 and 7. At the initial moment, temperature of the human body is 37°C and with time exposure it increases to a value close to 44°C. The greatest value of the temperature is inside the tumor but there are possible local maxima of temperature near the surface of the body. In order to avoid surface burns the human body would be surrounded with cold water bolus.

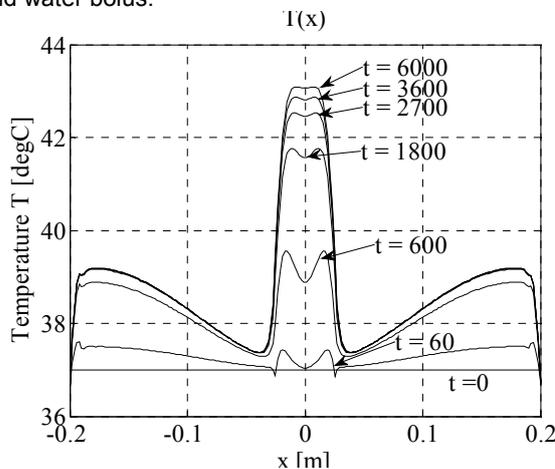

Fig.6. Temperature distribution along horizontal symmetry axis of the human body for different moments of time $t$ [s]

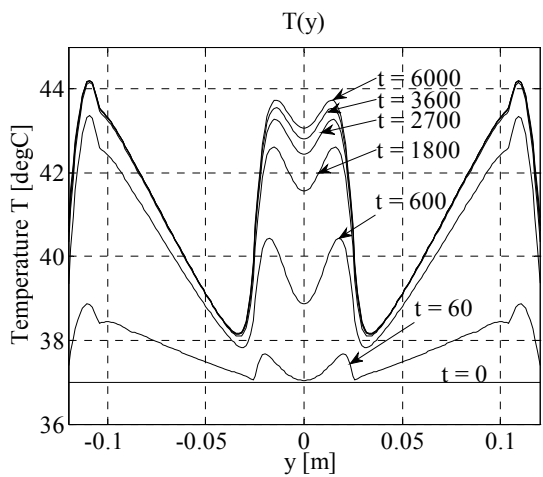

Fig.7. Temperature distribution along vertical symmetry axis of the human body for different moments of time $t$ [s]

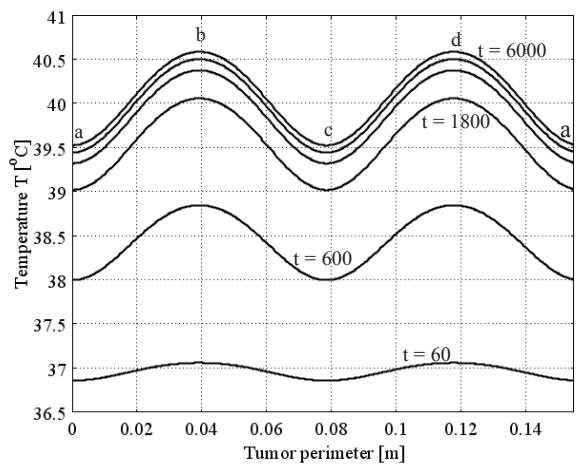

Fig.8. Temperature distribution along tumor perimeter for different points of the tumor

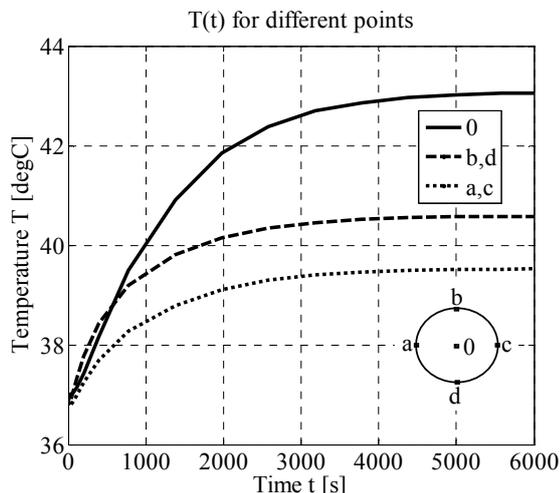

Fig.9. Time dependence of the temperature for different points of the tumor [points on the surface of the tumor: $a$ (-0.2, 0), $b$ (0, 0.12), $c$ (0.2, 0), $d$ (0, -0.12), point in the centre of the tumor $O$ (0, 0)]

The last two graphs show the temperature on the surface of the tumor. Temperature distribution along tumor perimeter for different points of the tumor is presented in Fig. 8. As one can see the tumor surface temperature is not constant and varies in the range from 39°C. to 40.5°C in steady state. However, the highest temperature occurs in the centre of the tumor as shown in Fig. 9.

**Summary**

Regional RF hyperthermia is an effective way of heating malignant tumors. Positive therapeutic effects of this treatment depends on the applied temperature, exposure time and the volume of the tissue exposed to electromagnetic fields. The effectiveness of heat treatment can be significantly increased by combining hyperthermia with other cancer treatments such as radiotherapy, chemotherapy, immunotherapy and gene therapy.
Numerical methods are often and readily used for dosimetric calculations for a number of important bioelectromagnetic issues. Thermal analysis of this problem by using the finite element method allows the estimation of influence of various model parameters on temperature growth in the specified area.
The analyzed model assumes the time-independence of the electromagnetic and thermal issues. A simulation showed that the electromagnetic and thermal processes have different orders. While the electromagnetic field reaches a steady state within a few microseconds, the temperature inside the tumor tissue may take even tens of minutes to stabilize. Therefore, hyperthermia treatment occupies about an hour and it is usually done once or twice a week. Hyperthermia has found application in the treatment of breast, skin, head and neck, brain, oesophagus, prostate, cervix of uterus, and urinary bladder tumors.

*Authors*: mgr inż. Piotr Gas, AGH University of Science and Technology, Department of Electrical and Power Engineering, al. Mickiewicza 30, 30-059 Krakow, E-mail: piotr.gas@agh.edu.pl